\documentstyle[12pt]{article}

\begin{document}

\begin{flushright} {\it December 22,1994 } \\
\end{flushright}
\begin{center}

{\bf
CHARGED PARTICLE PSEUDORAPIDITY DISTRIBUTIONS
IN Au+Al, Cu, Au, and U COLLISIONS AT~10.8~A$\cdot$GeV/c
}\\
\bigskip
\medskip
E877 Collaboration
\end{center}

\normalsize
\medskip
\begin{center}
J.~Barrette$^4$, R.~Bellwied$^{8}$, S.~Bennett$^8$, P.~Braun-Munzinger$^6$,
W.~E.~Cleland$^5$, M.~Clemen$^5$, J.~D.~Cole$^3$, T.~M.~Cormier$^8$,
G.~David$^{1}$, J.~Dee$^6$, O.~Dietzsch$^7$, M.~W.~Drigert$^3$,
J.~R.~Hall$^{8}$, T.~K.~Hemmick$^{6}$,
N.~Herrmann$^{2}$, B.~Hong$^6$, Y.~Kwon$^6$,
R.~Lacasse$^4$, A.~Lukaszew$^8$, Q.~Li$^8$, T.~W.~Ludlam$^1$,
S.~K.~Mark$^4$, S.~McCorkle$^1$, R.~Matheus$^8$,
J.~T.~Murgatroyd$^8$,
E.~O'Brien$^1$, S.~Panitkin$^6$, T.~Piazza$^6$, C.~Pruneau$^{8}$,
M.~N.~Rao$^6$, M.~Rosati$^4$, N.~C.~daSilva$^7$, S.~Sedykh$^6$,
U.~Sonnadara$^5$, J.~Stachel$^6$, N.~Starinsky$^4$, E.~M.~Takagui$^7$,
S.~Voloshin$^{5,}\footnote{On leave from Moscow Engineering Physics Institute,
     Moscow, 115409,  Russia}
$, G.~Wang$^4$, J.~P.~Wessels$^6$,
C.~L.~Woody$^1$, N.~Xu$^6$, Y.~Zhang$^6$, C.~Zou$^6$
\end{center}

\medskip
\begin{tabbing}
\hspace{1cm}\=                  \kill
\> 1. Brookhaven National Laboratory, Upton, NY 11973\\
\> 2. Gesellchaft f\"{u}r Schwerionenforschung, Darmstadt, Germany\\
\> 3. Idaho National Engineering Laboratory, Idaho Falls, ID 83415\\
\> 4. McGill University, Montreal, Canada\\
\> 5. University of Pittsburgh, Pittsburgh, PA 15260\\
\> 6. State University of New York, Stony Brook, NY 11794\\
\> 7. Universidade de S\~ao Paulo, S\~ao Paulo, Brazil\\
\> 8. Wayne State University, Detroit, MI 48202
\end{tabbing}

{\footnotesize
\centerline{ABSTRACT}
\begin{quotation}
\vspace{-0.10in}
%
We present the  results of an analysis of charged particle pseudorapidity
distributions in the central region in collisions of a Au
projectile with Al, Cu, Au, and U targets
at an incident energy of 10.8~GeV/c per nucleon.
The pseudorapidity distributions are presented as a function of
transverse energy produced in the target or central pseudorapidity regions.
The correlation between charged multiplicity and transverse energy
measured in the central region, as well as the target and projectile regions
is also presented.
We give results for transverse energy per charged particle as
a function of pseudorapidity and centrality.
\end{quotation}}

\bigskip

PACS number(s): 13.85.-t, 25.75.+r

\section{Introduction}

Global observables such as (pseudo-)rapidity particle density contain
valuable information on  the reaction dynamics and, indirectly,
on the degree of thermalization as well as the energy and entropy
densities reached in relativistic nucleus-nucleus collisions.
With light projectiles, pseudorapidity distributions have
been studied in detail previously both at AGS and CERN energies
(for a review, see Ref.~[1]).
Our collaboration earlier reported the measurements of charged particle
distributions [\ref{l814}], energy flow and stopping [\ref{ltcal1}],
transverse energy distributions [\ref{l877et}], in the collisions of
a Si beam with Al, Cu, and Pb targets.
Large energy deposition has been inferred from these measurements.
Extrapolating these results to heavy projectiles raises expectations to create,
in these collisions, the deconfined phase of quarks and gluons.
We report here measurements of the charged particle pseudorapidity
distributions in collisions of 10.8$\cdot A$~GeV/c Au beams with several
nuclear targets carried out in Experiment 877 operating
at the Brookhaven National Laboratory Alternating
Gradient Synchrotron (AGS).
We combine the multiplicity data with our earlier measurements~[\ref{l877et}]
of transverse energy ($E_t$) pseudorapidity distributions in Au+Au
collisions at the same energy to study the $E_t$ per charged particle as
a function of pseudorapidity and the centrality of the collision.
\section{Experimental setup}

The E877 experimental setup is shown schematically in Fig.~\ref{fe877}.
For the pseudorapidity distribution analysis we use data primarily from the
Multiplicity Detector, complemented by
data from the Participant and Target Calorimeters
(see the insert in Fig.~\ref{fe877}), which provide a measurement of
the centrality of a collision. ``Zero-degree'' energy (deposited mainly
by projectile spectators) is measured by the Uranium Calorimeter,
situated in the forward spectrometer.
The horizontal position of the beam particle
is measured by a pair of silicon strip detectors,
the  Beam Vertex Detectors,  shown in
the insert in Fig.~\ref{fe877} (BVer 1 and BVer 2). The
information from these detectors  is used on an event by event basis.
The mean vertical displacement of the incoming
 beam particle is estimated using
the information on the distribution of the ``centroid of hits''
in the multiplicity counter (see details in Section \ref{analysis}).
The angular divergence of the beam ($\sim 1$~mr) is much smaller than the
bin widths in $\eta$ and $\phi$ used for multiplicity distribution
analysis.
Most of upstream interactions are effectively rejected using
the pulse height from a 100~$\mu$m thick  Si surface barrier detector located
just upstream of the target.

The Multiplicity Detector, shown in Fig.~\ref{fdetector}, consists
of two identical silicon pad detectors, each of which
was made from a disc of silicon 300~$\mu$m thick and approximately
 3.8~cm in radius.
To reduce the number of $\delta$-electrons reaching these detectors,
 two 3~mm thick aluminum absorber plates
were placed upstream of each plane of silicon.
The active region of each detector is a ring of inner radius
1.4~cm and outer radius 3.4~cm divided into 512 pads.
The detectors are segmented into 8 concentric rings of 64 pads each.
One detector, located 3.37 cm from the target, covers
the pseudorapidity region $0.87<\eta<1.61$, and the other,
located 8.17 cm from  the target, covers the region $1.61<\eta<2.46$.
These values of pseudorapidity coverage correspond to the case where
the beam particle is incident at the center of the detector.
Due to the finite size of the beam spot and variation of the beam
position during the AGS spill,
the actual pseudorapidity  coverage is slightly larger.
The size of the pads corresponds approximately
to 0.1 in both $\eta$ and  azimuthal angle $\phi$,
which determines the angular resolution in this measurement.
Signals from the pads, after preamplification and shaping, are
sampled at the peak and digitized.
For the most central events the mean occupancy in pads
which see the highest track density is close to 0.3.

The Participant Calorimeter (PCal) [\ref{lpcal}]
is a lead/iron/scintillator sampling calorimeter.
It has a depth of four interaction lengths
and a radius of approximately 84 cm.
It is approximately azimuthally symmetric, built with four
identical quadrants.
Each quadrant of the PCal is divided into four azimuthal slices of 22.5$^{o}$.
Each slice is  divided radially into eight towers.
Longitudinally, the calorimeter
is divided into two electromagnetic depth segments and
two hadronic depth segments.
This division leads to  a total of 16$\times 8 = 128$ towers for
each quadrant and 512 towers for the entire calorimeter.
PCal  measures energy flow into the
polar angle region which corresponds to
pseudorapidity range 0.83 $<$ $\eta$ $<$ 4.7.
The Target Calorimeter (TCal) is made of 992 NaI crystals each 5.3 radiation
length deep.
It covers the backward hemisphere, corresponding to
the pseudorapidity range $-0.5\,<\,\eta \,<\, 0.8$.
For more details on TCal and the analysis of TCal data
see [\ref{ltcal1},\ref{ltcal2}].
The Uranium Calorimeter (UCal) consists of 25 modules and measures
the energy of particles entering the forward spectrometer through
a collimator with an opening of -115~mr$< \theta_x <$14~mr and
 -21~mr$< \theta_y <$21~mr.

The data were taken with several targets,
Al (242~mg/cm$^2$, approximately 1.9\% of an interaction length
for a gold projectile),
Cu (500~mg/cm$^2$, $\approx$2\%),
Au (540~mg/cm$^2$ and 980~mg/cm$^2$, $\approx$1\% and 1.8\%),
and U (575~mg/cm$^2$ and 1150~mg/cm$^2$, $\approx$1\% and 2\%).

\section{ Analysis}
\label{analysis}

\subsection{Pulse height spectra}

The pulse height distributions in the Multiplicity Detector
were first corrected for pedestal offsets
and differences in gain, and the non-functional channels were identified.
A channel was defined as good if the pulse height distribution
showed a minimum ionizing particle (m.i.p.) peak well
separated from the pedestal
(as in the example shown in Fig.~\ref{fldfit}).
Dead or noisy channels (altogether about 25\%) were removed in the analysis.
The corrected pulse height distributions were studied in a variety of ways.
In the vicinity of pedestals the distributions were fitted by a Gaussian;
this fit gives
the width of the electronic noise distribution in each particular channel, and
permits to evaluate the mean occupancy of the pad
($1-p_0$, where $p_0$ is the probability of the pad not being occupied).
The part of the distributions above the pedestals were fitted by
a Landau distribution convoluted with a  Gaussian describing
the  electronic noise and taking into account the effects of
atomic binding of the electrons [\ref{lh1}]
(this fit is similar to the analysis done in [\ref{lhelios}]):
\begin{equation}
f(\Delta)=\frac{1}{\sigma \sqrt{2\pi}}
\int_{0}^{\infty} f_{L}(\epsilon) \exp (\frac{-(\Delta-\epsilon)^2}
   {2\sigma^2})d\epsilon
\end{equation}
\begin{equation}
\sigma=(\delta_2+\sigma^2_{noise})^{1/2};
\end{equation}
Here $\Delta$ is the actual energy loss,
the variance $\delta_2$ is related to the effect of electron
atomic binding, and $\sigma^2_{noise}$ is the variance of electronic noise;
$f_{L}(\epsilon)$ is the Landau distribution function:
\begin{equation}
f_L(\epsilon)=\frac{1}{\xi}\Phi(\lambda);
\,\,\, \Phi(\lambda)=\frac{1}{2\pi i} \int_{c-i\infty}^{c+i\infty}
    \exp (u+\ln u +\lambda u)du
\end{equation}
\begin{equation}
\lambda=\frac{1}{\xi}(\epsilon - (\epsilon_{mp}-\xi \lambda_0))
=\frac{\epsilon- \epsilon_{mp}}{\xi}+\lambda_0;\;  \lambda_0=-0.225,
\end{equation}
where $\xi$ is the width of the distribution, and $\epsilon_{mp}$ is
the most probable energy loss; $c$ is an arbitrary real positive constant.
For the case of multiple ($n$) hits the parameters  $\xi$ and $\epsilon_{mp}$
are to be replaced by:
\begin{equation}
\xi_n=n \xi
\end{equation}
\begin{equation}
\epsilon_{n,mp}=n(\epsilon_{mp}+\xi \ln(n))
\end{equation}

To investigate the distribution in number of hits for the purpose of
evaluation of mean pad multiplicity,
two different fits to the pulse hight distributions were carried out.
The first fit assumed a  Poisson distribution in the number of
hits in the pad, and the parameter extracted was the mean occupancy,
which can be compared with the value extracted from the fit
to the electronic noise distribution.
The quality of the fit can be seen in Fig.~\ref{fldfit}, where the fitted
curves are shown along with the data from one of the pads.

In the second fit, the probabilities of single or double hits are
free parameters.
We find that the probability of double hits defined independently
exceeded the value expected from Poisson statistics.
For pads with mean occupancy $\approx 0.3$ the ratio of observed
double hits to the calculated value from Poisson statistics
is $\approx 1.3$.
This effect is understood to be the effect of $\gamma$-conversions
in the target and in the absorber.
The Monte Carlo simulation (described below) shows that
for a heavy target (Au or U) about 5\% of all hits are due to $\gamma$'s
from $\pi^0$ decays converting into $e^+e^-$-pairs.
Due to the small opening angle about 30\% of produced  $e^+e^-$-pairs
occupy the same pad of the multiplicity detector,  introducing
a non-Poissonian element in the distribution.
The correction for $\gamma$-conversions used in the analysis which takes
this effect into account is described below.

After the calibration was done and the procedure for the calculation of
the mean pad multiplicity was established, a ``hit''  threshold was
introduced corresponding approximately to one-half of
the peak of the minimum ionizing particle signal.
By varying the threshold position and comparing the results with
the information from the fit to the electronic noise distribution,
it was found that the occupancy can be defined in this
way with an accuracy better than 2\%.

\subsection{Corrections and Selection Criteria}

 {\em Event selection}.
In order for an event to contribute to the multiplicity analysis,
it is necessary to obtain the horizontal position of the beam particle.
Thus events with missing or ambiguous information from the vertex detector
were rejected.
Also, it is important to reduce the background from interactions
upstream of the target as much as possible.
Therefore, we reject events in which the pulse height in
the upstream silicon detector is below a threshold value close to
the energy loss peak of Au ions.

 {\it Beam position}.
To calculate the pseudorapidity corresponding to each pad the knowledge
of the  position of the interaction
point relative to the multiplicity detector is very important.
As mentioned above, the horizontal position is measured
for each event in the Beam Vertex Detector.
The following technique was used to define the relative position of the
Beam Vertex Detector with respect to the Multiplicity Detector.
We exploit the fact that
the hit centroid distribution is expected to be axially symmetric when the
beam particle is incident at the center of the Multiplicity Detector.
In each event the horizontal and vertical components of the  hit centroid
position are given by
\begin{equation}
C_H=\sum_i \cos {\phi_i};\,\,C_V=\sum_i \sin {\phi_i},
\end{equation}
where the sum is over all hits and $\phi_i$
is the azimuthal angle of the pad containing the $i$-th hit.
To avoid a bias in this part of the analysis from   dead pads,
a symmetrized dead pad mask was used, declaring  some good pads as  dead
to make the distribution of dead pads symmetric.
After the  coordinates of the centroid are calculated,
the horizontal component is plotted against the position provided by
the Beam Vertex Detector.
The results of this analysis are shown in Fig.~\ref{fbmcent}.
The beam position from BVer corresponding to $C_H=0$ gives
the the relative BVer and multiplicity detector
displacement (about 0.75~mm from Fig.~\ref{fbmcent}).
To understand the statistical fluctuations in the
plot near $x=0$, note that the most probable beam position from
 the BVer is about
$-4$~mm; thus the beam position near zero is very rare.

The average value of $C_V$ gives  information on the mean
 vertical position of the beam for any event sample.
A plot of $C_H$ versus the horizontal coordinate obtained
from BVer fixes the scale factor between the hit  centroid
position and the beam displacement.
For the data shown in  Fig.~\ref{fbmcent} this scale factor
is 4.1~mm, which gives us about 1.2~mm for the mean vertical beam offset.
One can also see from the plot that there is no correlation between
vertical and horizontal components of the beam position.

The correction for multiple hits was done pad by pad using the average
occupancy.
Since the mean pad occupancy depends on beam position,
the correction was done separately for each value of this variable.
For this purpose the corresponding pseudorapidity, azimuthal
angle, and solid angle in these variables  were calculated for every pad
as a function of both vertical and horizontal beam position.
The Multiplicity Detector consists of two silicon pad detectors located
at different distances from the target and covering
different pseudorapidity regions.
For non-zero beam positions these regions overlap.
It was verified that the distributions calculated using the data from
different detectors coincide in the overlapping region for the values
of the beam mean vertical displacement and the Multiplicity Detector
relative offset used in the analysis.
The other independent check is that the resulting azimuthal angular
distributions are flat in different pseudorapidity windows.
These methods are sensitive to displacements at the level
of 0.2--0.3~mm.

 {\it Upstream interactions}.
Although most of the upstream interaction events are effectively
rejected using the information from the upstream Si detector
and the Beam Vertex Detectors, it is important for collisions
of medium centrality  and for data taken with  light targets
to perform a background subtraction to correct for residual
upstream interactions.
The subtraction was done using the data taken with empty target frame.
The relative contamination by upstream interactions is different for
events triggered by TCal or PCal.
Almost all upstream interactions result in relatively low TCal $E_t$, but
PCal $E_t$ can be rather significant.
For heavy targets the admixture of upstream interactions in
the event sample is negligible for events with
TCal $E_t>10$~GeV, and only about 20\% for the region $E_t \approx $4--5~GeV
(the corresponding differential cross section $d\sigma /dE_t$ is presented
in the following section in Fig.~\ref{fdet1}).
For the Al target the admixture is about 40\%
in the same region (TCal $E_t \approx $4--5~GeV).
In contrast to the heavy targets the admixture of upstream interactions
for the Al target increases with TCal $E_t$
because of the sharp drop of $d\sigma /dE_t$ for interactions in the target;
for events with TCal $E_t>7$~GeV, upstream interactions become dominant.
For the Au target the admixture of upstream interactions in PCal $E_t$
regions centered at 65~GeV, 130~GeV, and 190~GeV is about
70\%, 40\%, and 10\% respectively.

{\it Delta electrons}.
One of the most important corrections is the subtraction of hits due to
$\delta$-electrons produced in the target.
The aluminum absorber located in front of each detector plane
reduces the number of $\delta$-electrons by about a factor of 10, but
nevertheless  their contribution is not negligible.
It is not possible to extract the pseudorapidity distributions of
$\delta$-electrons (needed for the correction) directly from the data.
The data for ``beam'' events (no interaction in the target) provide the
number of produced  $\delta$-electrons and their pseudorapidity
distribution for the case when the incoming nucleus traverses
the entire length of the target.
Unfortunately, this distribution cannot be used for the correction
for normal events, because the path length of the
incoming nucleus in the target is different from that of a beam track,
and thus  the multiple scattering and absorption effects are also different.
To understand the effect of the absorbers on the $\delta$-ray
energy and angular distribution and to calculate the pseudorapidity
distribution needed for the correction a detailed
simulation was performed using the GEANT (version 3.16) package.

The GEANT results were checked by comparison with the data in several ways.
The total number of produced $\delta$-electrons
in ``beam'' (no interaction in the target)
events and their pseudorapidity distribution were compared directly
with the data for different targets and target thicknesses, and
good agreement was found.
For data with interactions in the target we compare the total number of
$\delta$-electrons seen by the Multiplicity Detector.
We obtain this value from the data by extrapolation of the number of
hits in the Multiplicity Detector $N_{ch}^{raw}(E_t)$
as a function of $E_t$ to the point $E_t=0$ (Fig.~\ref{fnch_vs_et})
from the region $E_t > 4$~GeV, where $N_{ch}^{raw}$ grows linearly with $E_t$
and the number of produced $\delta$-electrons does not depend on $E_t$.
The extrapolation gives for the number of $\delta$-electrons
values of about 13 and 24 for the 1\% and 2\% Au targets
respectively, which agree well with the GEANT simulation.
The number of produced $\delta$-electrons does not depend on $E_t$
in the region $E_t >4$~GeV because $\delta$-electrons are mostly produced
by the projectile nuclei before the collision due to the relatively
large value of projectile charge $Z$, whereas the production of
$\delta$-electrons by final state particles is negligible.
For very low $E_t$ events one expects an increase of
$\delta$-electron contribution to $N_{ch}^{raw}$ due to
projectile fragmentation into high $Z$ fragments, as
can be seen in Fig.~\ref{fnch_vs_et}.
The value of $N_{ch}^{raw}$ at $E_t=0$ corresponds to the number of
$\delta$-electrons produced by the projectile traversing the entire target.

The GEANT $\delta$-electron simulations
have no free parameters which affect the results.
The cutoff parameters in GEANT affect the low energy part of
the $\delta$-electron energy spectrum, but because of the absorption
of low energy electrons in the aluminum
absorber plates, we are insensitive to the choice of these parameters.
The pseudorapidity distributions of $\delta$-electrons used for
the corrections for 1\% Au target data are shown in Fig.~\ref{fdn_cor}
(curves 2 and 3).
At the level of accuracy required for this analysis,
he distribution of $\delta$-electrons does not depend on the centrality
of the collision.

 {\it Gamma conversions}.
Another important correction is due to $\gamma$-conversions in the target
and absorber. This correction was calculated using
FRITIOF~[\ref{lfritiof}]  and RQMD~[\ref{lrqmd}] generated events
combined with a GEANT simulation of the detector.
In particular  the ratio of pseudorapidity distributions
of hits in the multiplicity detector due to $\gamma$-conversions and
the pseudorapidity distribution of charged particles was calculated
for different centralities and target thicknesses.
The results do not depend on the particular event generator used.
They show that
for a 1\% Au target approximately 5\% of the charged multiplicity
seen in the multiplicity detector is due to $\gamma$-conversions in
the target and an additional 1\% is due to conversions in the absorber.
The effect depends slightly on the pseudorapidity: it is about 8\%
in the low part of the pseudorapidity region
and about 4\% for high pseudorapidities.
About 30\% of all $e^+e^-$-pairs  occupy the same pad of
the Multiplicity Detector.
This effect results in the distortion of the Poisson statistics
for the hit multiplicity distribution.
The distortion caused by the two-particle correlations for produced
secondaries is negligible, due to the very small magnitude of the
correlations~[\ref{l814cor}].
The method used to correct for the $\gamma$-conversions
is to apply the Poisson correction to the value
calculated from the  mean pad occupancy and then to subtract
the distribution of electrons and positrons considering
the pairs occupying the same pad as one charged particle.
For the details of this procedure see Appendix A.
The distribution used for the correction of
the  1\% Au target data is presented in Fig.~\ref{fdn_cor} (curve 1)
and has been used to obtain the charged particle multiplicity
distribution for the highest centrality bin (see Fig.~\ref{fdn}).
The correction for other values of centrality
scales approximately as $dN_{ch}/d\eta$.

\section{ Results}
\label{sresults}

\subsection{ Correlations among global variables}

The charged particle multiplicity measured in the Multiplicity Detector
is strongly correlated with transverse energy deposited in TCal
and/or PCal, and anti-correlated to the (``zero-degree'') energy
deposited in UCal.
We present these correlations in Fig.~\ref{fnraw_corr}, where we plot
the uncorrected (raw) number of hits in the Multiplicity Detector
versus energy in each of the detectors.
The relative abundance of the events with high multiplicity is an effect
of trigger thresholds.
The shapes of the the distributions in TCal/PCal $E_t$ for
the events with fixed multiplicity, and the distributions in
multiplicity for the events with fixed TCal/PCal $E_t$, are close
to Gaussian distributions.
The length  of the crosses in Fig.~\ref{fnraw_corr} indicate the widths
of each distribution in different regions of the plot.

The transverse energy deposited in the target (and/or
central) region is strongly anticorrelated with the impact
parameter of the collision.
Therefore one can infer from Fig.~\ref{fnraw_corr} that
there are no drastic changes in the charged multiplicity fluctuations
between central and non-central events.
The width of the correlation between charged particle multiplicity and
the PCal $E_t$, which corresponds to the transverse energy deposited by all
(charged and neutral) particles in the same pseudorapidity region,
indicates that there are no large scale
fluctuations in the ratio of charged and neutral particle multiplicities.

For the current analysis the TCal $E_t$ data were
used primarily as the measure of centrality of the collisions.
Note that these data are almost free from contamination
by upstream interactions (except for low centrality events,
where the correction for upstream interactions is small but not negligible).
To compare our present multiplicity data with the published $dE_t/d\eta$
measurements in Au+Au collisions~[\ref{l877et}],
we use as the measure of centrality PCal $E_t$ data,
which were effectively corrected for the leakage in the calorimeter
through normalization to our previous measurements [\ref{l877et}].

The differential cross sections in TCal $E_t$ for different targets are shown
in  Fig.~\ref{fdet1}.
The centrality of the collisions can be inferred using the ratio
$\sigma_{top}(E_t)/\sigma_{geom}$ (shown on bottom plate of Fig.~\ref{fdet1}),
where $\sigma_{top}(E_t)$ is defined as
\begin{equation}
\sigma_{top}(E_t)=\int_{E_t}^{\infty}d\sigma /dE_t dE_t,
\label{etop}
\end{equation}
and the geometrical cross section for the collision of $A$ and $B$ nuclei
is $\sigma_{geom}=\pi (R_A+R_B)^2;\, R_{A,B}=1.2 A^{1/3}$~fm.
The U, Cu, and Al targets data were obtained by (multilevel) triggering on
PCal $E_t$.
The width of the correlation between PCal $E_t$ and TCal $E_t$ is relatively
large in comparison with the difference between different trigger thresholds.
A special weighting technique was used to measure  $d\sigma/dE_t({\rm TCal})$
from data triggered with  PCal $E_t$ at several thresholds (see Appendix B).

The differential cross sections $d\sigma/dE_t$ (TCal, PCal)
for the Au target agree well with our previous measurements~[\ref{l877et}].
Central collisions of Au nuclei with a U target produce about 20\%
more transverse energy in the TCal region than collisions with a Au target.
However in the more central (PCal) region the produced transverse
energy differs only by about 5\%.
We observe a similar difference between the two targets in the peak values of
charged particle densities (see below).

\subsection{ Charged particle pseudorapidity distributions}

Charged particle pseudorapidity distributions for different targets
are presented in Fig.~\ref{fdn} for different centralities.
The centrality for the different TCal $E_t$ regions can be estimated from the
ratio
$\sigma_{top}(E_t)/\sigma_{geom}$  shown in  Fig.~\ref{fdet1}.
For the discussion below it is important to note that for values of $E_t$
close to its maximum value the mean multiplicity depends weakly
on the $E_t$ cut (for Au+Au collisions this is clearly evident
in Figures~\ref{fnch_vs_et} and~\ref{fnraw_corr} for TCal $E_t$
greater than 21~GeV).
This suggests a simple criterion for defining the similar centralities
for collisions between different projectile and target nuclei.
The highest $E_t$ bins shown in Figures~\ref{fdn}(a)--(c) are chosen
to satisfy the requirements of a weak dependence of the multiplicity on
the $E_t$ cut. These cuts correspond to values of
$\sigma_{top}(E_t)/\sigma_{geom}$ from approximately 2\% for
Au+Al to 0.2\% for Au+U, which correspond closely to cuts used to define
central events in our studies of transverse energy
production~[\ref{l877et}].

We have studied the effects of systematic errors in
the charged particle pseudorapidity distributions as a function of
pseudorapidity.
{}From uncertainties in the corrections to the data
(beam position, $\delta$-rays, $\gamma$-conversions) we estimate
the systematic error in the magnitude to be  about 3\% in the
mid-pseudorapidity region for central Au+Au collisions.
The uncertainty is slightly larger in the low pseudorapidity region
(about 5\%) and for lower centralities.
We have also considered possible errors in the position of the peak
found from Gaussian fits to the pseudorapidity distributions.
These errors were evaluated by a variation of parameters in the
distributions used for the corrections.
It was found that the centroid of Gaussian fit is surprisingly
stable for such variations.
The systematic uncertainty
was estimated to be not more than 0.03 units of pseudorapidity;
it is mainly due to the uncertainty in the vertical position of the beam
(which gives a contribution of about 0.02 units of pseudorapidity)
and  to the uncertainty in the correction
for $\gamma$-conversions (also about 0.02 units).
The uncertainties due to statistical errors, are much smaller, than
the systematic uncertainties.

For central and mid-central collisions and heavy targets (Au and U)
the pseudorapidity distributions are well fitted by Gaussians.
We show the fits for the highest $E_t$ regions in
Fig.~\ref{fdn}~(a) and Fig.~\ref{fdn}~(b).
If we compare the fits for both targets, we see that for the U target
the height is about 6\% larger (about 285 and 268, respectively).
The position of the peak is shifted to a lower
value of pseudorapidity (1.71 for U, and 1.76 for the Au target), consistent
with naive expectations for the heavier target.
The distribution has the same width (about 1.05) as for the Au target.
Note that the widths of analogous distributions
for very central collisions of a Si beam with Al, Cu, and Pb
targets~[\ref{l814}] have very nearly the same value.
For gold collisions with Cu and Al targets (Fig.~\ref{fdn}~(c))
the shape of the distributions becomes non-Gaussian.
The peaks of the distributions are shifted to the higher values of
pseudorapidity, as expected for collisions with
light targets.

For low centrality $E_t$ regions the pseudorapidity distributions
for Au+Au and Au+U collisions exhibit an enhancement for
low pseudorapidities.
Monte Carlo studies imply that this asymmetry is caused by slow protons
from target fragmentation.
This hypothesis was checked by studying separately pseudorapidity
distributions of hits with pulse heights around the minimum ionizing peak
(where one does not expect contributions from slow protons).
The distributions of low pulse height  hits do not show such an
asymmetry; thus the low pseudorapidity enhancement appears to be
entirely due to hits with high  pulse heights.

We compare our results with the RQMD [\ref{lrqmd}]
and FRITIOF [\ref{lfritiof}] event generator predictions
in Fig.~\ref{fdn_mc} for the most central Au+Au collisions.
The centrality for Monte Carlo events was defined in accordance with
the top cross section calculated using the transverse energy
deposited in the TCal or PCal pseudorapidity region,
as is done for the data.
When using TCal $E_t$ as a measure of centrality
both event generators underpredict the peak value of the
pseudorapidity density (Fig.~\ref{fdn_mc}~(a)).
FRITIOF also overpredicts the position of the peak
($\approx 2.2$ in comparison to the experimental value of 1.76).
Similar trends were seen for FRITIOF results
on pseudorapidity distributions of transverse energy~[\ref{l877et}].
In Fig.~\ref{fdn_mc}~(b) we compare the event generator predictions
with the data using PCal $E_t$ as a measure of centrality.
Whereas the difference between FRITIOF results and the data remains almost
the same as in Fig.~\ref{fdn_mc}~(a),
the agreement between RQMD and data is significantly better.
The origin of the different behavior is in the difference
in the correlations between $N_{ch}$ with TCal $E_t$ and PCal $E_t$
for the data and the event generators.
The FRITIOF $N_{ch}$ -- TCal $E_t$ and $N_{ch}$ -- PCal $E_t$ correlations
are close to the observed ones, and the results of comparison of
pseudorapidity distributions is insensitive to the choice of
 TCal or PCal $E_t$ as a measure of centrality.
On the other hand RQMD exhibits a very tight
  $N_{ch}$ -- PCal $E_t$ correlation, while
the $N_{ch}$ -- TCal $E_t$ correlation is rather loose.
This results in lower mean $N_{ch}$ for the events with highest TCal $E_t$
in comparison with the mean value of $N_{ch}$ for events
with highest  PCal $E_t$.
Thus, for the same value of top cross section,
the charged particle pseudorapidity densities for RQMD are quite
different for the two cases.

\subsection{ Comparison with $dE_t/d\eta$ and evaluation of $E_t$
per charged particle}

Below we combine multiplicity data with our earlier measurements
[\ref{l877et}] of $dE_t/d\eta$ in Au+Au collisions at a similar,
but not identical, energy.
Due to the strong correlation between impact parameter and transverse energy
produced in the central region (in our case PCal $E_t$)
it was found useful [\ref{l877et}] to introduce the value
$E_t^0$, the $E_t$ for an average collision with impact parameter
$b<0.5$~fm ($\sigma_{top}(E_t^0)/\sigma_{geom}=0.22\%$), and
use the ratio $E_t/E_t^0$ as a measure of centrality.
For our case this yields $E_t^0\approx 318$~GeV.
In Fig.~\ref{fdn_au_p}(a) we present the charged particle pseudorapidity
distributions for different PCal $E_t$ regions centered approximately
at $E_t/E_t^0=0.2,\,0.4,\,0.6,\,0.8,$ and 1.0.
The smooth curves represent Gaussian fits to the data.
The peak position~($\eta_{peak}$) and the width~($\sigma_{eta}$) of
Gaussian fits are presented in Fig.~\ref{fdn_au_p}(b).
The open circles in Fig.~\ref{fdn_au_p}(b) show
the dependence of parameters of Gaussian fit to $dE_t/d\eta$
distributions from Ref.~[\ref{l877et}].
The peak position of $dN_{ch}/d\eta$ is very close to that for
$dE_t/d\eta$.
It is smaller for lower centrality, presumably due
to the relatively larger contribution of slow protons from the target.
The width of the $dN_{ch}/d\eta$ distribution decreases as centrality
increases (similar to $dN_{ch}/d\eta$ data for both the Au and
Si beams), and is larger than the corresponding width
of the $dE_t/d\eta$ distribution.

We use the $dE_t/d\eta$ distributions for the same centrality
(value of $E_t/E_t^0$) as shown in Fig.~\ref{fdn_au_p}
to calculate $E_t$ per charged
particle as a function of pseudorapidity and centrality,
 and to compare them with
RQMD and FRITIOF predictions (see Fig.~\ref{fetperpart}).
By  $E_t$ per charged particle we mean the ratio of
average value of total transverse energy to the average
charged particle multiplicity $E_t/N_{ch}$ (which is not the mean
transverse energy of charged particles).
Note that the value of transverse energy per charged particle
is less sensitive than the absolute pseudorapidity spectra
to uncertainties in defining the centrality cuts
for experimental and Monte Carlo generated events.
The pseudorapidity dependence of $E_t$ per charged particle shown in
 Fig.~\ref{fetperpart}(a) was calculated as a ratio of Gaussian fits to
$dE_t/d\eta$~[\ref{l877et}] and $dN_{ch}/d\eta$ distributions.
The observed large value of $E_t$ per charged particle
about 0.75~GeV (for central collisions and in the central
 pseudorapidity region) is rather remarkable.
If we take into account that the $dE_t/d\eta$ spectra were measured
at slightly higher energy (11.4~GeV/nucleon compared to 10.8~GeV/nucleon
for current data) and re-scale $E_t$ per charged particle
with the available energy [1], which differs for both cases by
approximately 4\%,  we get the value of 0.72~GeV.
This value is significantly higher than that in p+Pb
(about 0.45~GeV~[\ref{l814pa}]), Si+Al and Si+Pb collisions
(0.55--0.59~GeV and 0.52--0.54~GeV, respectively [\ref{l814},\ref{l877et}])
at an even higher beam energy of $\approx$14.6~GeV/nucleon.
Note that we compare the $E_t$ per charged particle
for the central events triggered on  $E_t$.
The pseudorapidity distributions in References~[\ref{l814}]
and~[\ref{lkith}] were studied as a function of total charged
multiplicity and cannot be used directly;
but using the correlation between multiplicity and PCal $E_t$
it is possible to estimate the peak value of the distribution
for collisions with the highest PCal $E_t$.
For example, for Si+Pb collisions this estimate gives a value of
approximately 115--120 particles
per unit of pseudorapidity for central Si+Pb collisions.
Combining this value with the peak value of $dE_T/d\eta \approx 62$~GeV
observed in [\ref{l877et}] one finds the value of $E_t$ per charged
particle cited above.

The comparison of model calculations with the data shows that
the RQMD event generator describes the pseudorapidity and
nontrivial centrality dependence of $E_t$ per charged particle rather well,
although the generator predicts the peak position to be
at a slightly higher value of pseudorapidity.
It cannot be excluded that this apparent disagreement is in part
due to the representation of  $dN_{ch}/d\eta$ and $dE_t/d\eta$ as
Gaussian functions and to the fact that the two data sets were taken
at slightly different beam energies.
FRITIOF does not reproduce the pseudorapidity dependence and,
in disagreement with the data, shows no dependence on centrality.

The possible origin of rather high value of $E_t$ per charged particle
was studied using the RQMD event generator, since it
exhibits a dependence on centrality and pseudorapidity which is quite
similar to the data .
For this purpose Si+Pb at 14.6~GeV/c and Au+Au at 11.4~GeV/c
collisions were studied.
It was found that centrality dependence is almost totally
due to changes in transverse energy deposited by nucleons.
For less central events (and the collisions of nuclei of very different sizes,
such as Si and Pb) the relative contributions of target spectators
is rather large, which causes a decrease in $E_t$ per charged particle.
The second reason for the difference in $E_t$ per charged particle between
central collisions of light and heavy projectiles is a difference in
the ratio of charged to all final state particles (caused by different
relative numbers of protons and neutrons).
It was observed in this model, that the transverse energy of
produced particles (pions, kaons) do not exhibit any strong dependence on
incident energy (within the energy range considered),
size of the target or projectile, and centrality of the collision.

\section{ Conclusion}

We have presented an analysis of charged particle multiplicity
distributions in collisions of Au projectiles with Au, U, Cu, and Al
targets with different centralities.
The results are corrected for beam movement, upstream
interactions, $\delta$-ray production, and $\gamma$-conversions.
In studying the correlation between charged multiplicity and energy
deposited in the calorimeters, we observe little or no change in
the fluctuations of charged multiplicity as a function
of centrality and no large scale fluctuations between charged
multiplicity and transverse energy deposited in the same
pseudorapidity region.
The maximum value of the charged particle pseudorapidity density for
very central
Au+Au collisions is close to 270 and about 5\% larger for the U target.
The transverse energy per charged particle grows with increasing
 centrality.
For central Au+Au collisions it is close to 0.72~GeV, significantly
higher than in p+Au or Si+Pb collisions.

The FRITIOF event generator underpredicts the number of produced
particles in the central region.
The peak of the FRITIOF pseudorapidity distribution occurs at a much larger
value of pseudorapidity than in the data.
RQMD also underpredicts the charged particle density if one selects
central events using transverse energy in the target region.
If transverse energy in the central pseudorapidity region is chosen
as a measure of the centrality, the description is better.
The reason for this lies in the looser correlation between multiplicity
and transverse energy in the target fragmentation region in the events
generated by RQMD, in comparison with the data.
The RQMD description of the centrality dependence of $E_t$ per
charged particle (calculated in the region $1.5<\eta < 2.0$)
is rather good; considering the pseudorapidity dependence of
$E_t$ per charged particle calculated for central collisions
RQMD predicts the peak position of the distribution
at slightly higher pseudorapidity value than data does.
It is likely that the degree of re-scattering is even more important
in the  multiparticle production in heavy nucleus collision than it is
implemented in the event generators;
thus the production of thermalized hadron matter
is more probable in such collisions.

\section*{ Acknowledgments}

This research was supported, in part, by the U.S. DOE, the NSF,
and Natural Sciences and Engineering Research Council of Canada.

\newpage
\section*{Appendix A: Correction for $\gamma$-conversion}

Let us denote by $n$ the mean charged particle multiplicity in
a particular pad; $n$ is the quantity of interest.
The $\gamma$-conversions to $e^+e^-$-pairs result in additional
hits in the pad characterized by the mean multiplicity of
uncorrelated (from different pairs) particles
$n_{\gamma 1}$ and mean number of pairs $n_{\gamma 2}$,
when both the $e^+$ and $e^-$ from the pair occupy the same pad.
In terms of these quantities the probability of a pad not being
occupied (the input for the current analysis) is:
\begin{equation}
p_0=1-e^{-n-n_{\gamma 1}-n_{\gamma 2}}.
\end{equation}
The corresponding (to Poisson statistics) effective mean multiplicity
equals:
\begin{equation}
\tilde{n}=-\ln (1-p_0)=n+n_{\gamma 1}+n_{\gamma 2}.
\end{equation}
To obtain the value $n$ one should subtract from
$\tilde{n}$ not the true mean number of $e^+$ and $e^-$
from $\gamma$-conversions ($n_e=n_{\gamma 1}+2 n_{\gamma 2}$),
but the number of ``apparent'' particles, considering the pairs
occupying the same pad as one particle ($n_{\gamma 1}+n_{\gamma 2}$).

The distortion to Poisson statistics for multiple hits in the same pad
due to $\gamma$-conversions can be
evaluated from the probabilities of single and double hits:

\begin{equation}
p_1=(n+n_{\gamma 1}) e^{-n-n_{\gamma 1}}(1-e^{-n_{\gamma 2}}),
\end{equation}
\begin{equation}
p_2=(n+n_{\gamma 1})^2 e^{-n-n_{\gamma 1}}(1-e^{-n_{\gamma 2}})/2 +
     n_{\gamma 2} e^{-n_{\gamma 2}}(1-e^{-n-n_{\gamma 1}})
\end{equation}

\section*{Appendix B: Reconstruction of distribution from multilevel triggered
data}

Below we discuss a technique which permits the construction
of unbiased experimental distributions from multilevel triggered data
for the case where the variable of interest is weakly correlated
with the variable on which the data are triggered.
We make use of the expression which gives the distribution $dw/dx$
in a quantity $x$, as an integral over the distribution $dw/dy$
in another quantity $y$,  on which the trigger
decision is based:
\begin{equation}
\frac{dw}{dx}=\int \frac{dw}{dy}(y) \frac{dP}{dx}(x;y) dy,
\label{eb1}
\end{equation}
where $dP/dx(x;y)$ is the distribution in $x$ for events at a fixed
value of $y$.  It is important to recognize that, whereas $dw/dy$ is the
distribution of the triggered quantity which is strongly influenced by
downscaling (i.e. only a known fraction of the events above
the trigger threshold are recorded), the function $dP/dx(x;y)$ expresses
the natural correlation
between the two quantities and is independent of the trigger.
If downscaling is introduced in the trigger,
the distribution in $y$ of the events which are written
to tape is given by:
\begin{equation}
\frac{d\tilde{w}}{dy}=W(y) \frac{dw}{dy},
\label{eb2}
\end{equation}
where $W(y)$ is the probability for the event with certain value of $y$
to be written onto the tape and is constant within each trigger level region.
We combine Eqs.~(\ref{eb1})-(\ref{eb2}) to obtain
\begin{equation}
\frac{dw}{dx}=\int \frac{1}{W(y)}
        \frac{d\tilde{w}}{dy}(y) \frac{dP}{dx}(x;y) dy,
\label{ef}
\end{equation}
which can be used directly for the calculation of distributions
of quantities from multilevel triggered data.
Eq.~(\ref{ef}) has a very simple interpretation:
in the calculation of distributions of some quantity $x$
each event should be weighted with the inverse probability
of the event being triggered.
In other words, to calculate the natural
distribution of events in $x$, we evaluate
\begin{equation}
\frac{\Delta N}{\Delta x}=\sum_{{\rm events} \; {\rm in}\; \Delta x}
\frac{1}{W(y)},
\end{equation}
where the sum is taken over all triggered events written on tape with value
of $x$ within the region $\Delta x$.

\newpage
\section*{ References}          
\begin{enumerate}

\vspace{-10pt}
\item \label{lsy}
        For a review of both CERN and AGS data, see
 J.~Stachel and G.~R.~Young, Annu. Rev. Nucl. Part Sci. {\bf 42}, 237 (1992),
  and references quoted there.

\vspace{-10pt}
\item \label{l814}
        E814 Collaboration, J. Barrette {\it et al.},
Phys.\, Rev. {\bf C46}, 312 (1992).

\vspace{-10pt}
\item \label{ltcal1}
        E814 Collaboration, J. Barrette {\it et al.},
Phys.\, Rev.\, Lett. {\bf 64}, 1219 (1990).

\vspace{-10pt}
\item \label{l877et}
        E814/E877 Collaboration, J. Barrette {\it et al.},
Phys.\, Rev.\, Lett. {\bf 70}, 2996 (1993)

\vspace{-10pt}
\item \label{lpcal}
        The details of PCal construction and calibration are given
in J.~Simon-Gillo {\it et al.}, Nucl. Instrum. Methods
{\bf A 309}, 427 (1991);
D.~Fox {\it et al.}, Nucl. Instrum. Methods
{\bf A 317}, 474 (1992);

\vspace{-10pt}
\item \label{ltcal2}
        E814 Collaboration, J. Barrette {\it et al.},
Phys.\, Rev. {\bf C45}, 819 (1994).

\vspace{-10pt}
\item \label{l814cor}
        E814 Collaboration, J. Barrette {\it et al.},
Phys.\, Rev. {\bf C49}, 1669 (1994).

\vspace{-10pt}
\item \label{lh1}
        S. Hancock {\it et al.}, Phys.\, Rev. {\bf A48}, 615 (1983);
S.~Hancock {\it et al.}, Nucl.\, Instrum.\, Methods
     {\bf B1}, 16 (1984).

\vspace{-10pt}
\item \label{lhelios}
        HELIOS Collaboration, T.~\AA kesson {\it et al.},
Nucl.\, Phys. {\bf B333}, 48 (1990).

\vspace{-10pt}
\item \label{lkith}
        K. Jayananda, PhD Thesis, University of Pittsburgh, 1991

\vspace{-10pt}
\item \label{lfritiof}
        B. Andersson, G. Gustafson, and B. Nilsson-Almqvist,
Nucl.\, Phys. {\bf B281}, 289 (1987); the parameters for the AGS energies
were taken from: J.~Costales, E802 Internal report No.~E-802-MEM-8,
1988 (unpublished).

\vspace{-10pt}
\item \label{lrqmd}
        H. Sorge, H. St\"{o}cker, and W. Greiner, Ann.\, Phys. (N.Y.)
{\bf 192}, 266 (1989).

\vspace{-10pt}
\item \label{l814pa}
        E814 Collaboration,
M. Rosati, Nucl.~Phys., {\bf A566}, 597c (1994);
J. Barrette {\it et al.}, ``Transverse energy
and charged multiplicity in p-nucleus collisions at 14.6~GeV/nucleon'',
E814/E877 preprint, 1994.

\end{enumerate}

%
%
%
%

\newpage
\section*{Figure Captions}

\begin{enumerate}

\item    \label{fe877}
Experimental setup of E877 at BNL. For this analysis, we use data
from the Multiplicity Detector, Target, and
Participant calorimeters (see insert).

\item    \label{fdetector}
E877 Multiplicity Detector, consisting of two identical
  300 $\mu$m  thick silicon disks, segmented
into 512 pads. The combined pseudorapidity coverage is  $0.87<\eta <2.46$.

\item    \label{fldfit}
Fits of the pulse height distributions by modified Landau distributions.
Contribution from 1, 2, and 3 m.i.p. are shown. The dashed line indicates
the Gaussian fit to the electronic noise.

\item    \label{fbmcent}
Distribution of the horizontal (solid histogram) and
vertical (dashed histogram) coordinates of
the hit centroid as a function of the beam position measured in the
 Beam  Vertex Detector.
Straight lines represent  linear fits to the distributions.

\item \label{fnch_vs_et}
The dependence of mean number of hits in the Multiplicity Detector
as a function of TCal $E_t$ is shown for 1\% (solid histogram) and
2\% (dashed histogram) Au targets.
The extrapolations to $E_t=0$ gives an estimate of the numbers
of $\delta$-electrons.

\item    \label{fdn_cor}
Charged particle pseudorapidity distributions used as  corrections
for Au+Au collisions (1\% target).
(1) Distribution due to $\gamma$-conversions for  TCal $E_t$ region
22--28~GeV, (2) distribution of $\delta$-rays produced in
the target and in the absorbers, and (3)Distribution of $\delta$-rays
from upstream material.

\item \label{fnraw_corr}
Transverse energy in  TCal, PCal, and downstream energy deposited in UCal
plotted {\it versus}  the raw number of hits in the Multiplicity Detector.
The crosses indicate the width ($\sigma$) of the distributions in each
variable.

\item    \label{fdet1}
(Top panel):
Differential cross sections in TCal  $E_t$ for collisions of
a Au projectile with U, Au, Cu, and Al targets.
For clarity  Au+Au, Au+Cu, and Au+Al cross sections are multiplied
by factors of 0.1, 0.01, and 0.001 respectively.
(Bottom panel): The cross section integrated above $E_t$
(see Eq.~\ref{etop} and text thereafter),
 normalized to the geometrical cross section.

\item \label{fdn}
 Charged particle pseudorapidity distributions for
different centralities (TCal $E_t$ regions).
For the most central region we show also the Gaussian fit to
the distribution.
(a) Au+Au collisions;
(b) Au+U collisions;
(c) Au+Cu and Au+Al collisions.

\item \label{fdn_mc}
 (1) Experimental charged particle pseudorapidity distribution
for Au+Au collisions,  compared  with RQMD (2)
and FRITIOF (3) predictions for the equivalent centrality;
(a) TCal $E_t$ energy region 22--28~GeV ;
(b) PCal $E_t$ energy region 310--330~GeV.

\item \label{fdn_au_p}
(a): Charged particle pseudorapidity distributions in Au+Au collisions for
the different centralities (PCal $E_t$ regions).
The curves are Gaussian fits to the data.
(b): Peak position and the width of the Gaussian fit to
$dN_{ch}/d\eta$ as a function of centrality
($E_t/E_t^0$, for details, see text) in comparison with parameters
of the analogous fits to $dE_t/d\eta$~[\ref{l877et}], shown as
open circles.

\item \label{fetperpart}
Transverse energy per charged particle for Au+Au collisions,
compared with predictions from RQMD and FRITIOF.
The data are derived from Gaussian fits to $dE_t/d\eta$ taken at
11.4~GeV/nucleon and $dN_{ch}/d\eta$ taken at 10.8~GeV/nucleon,
as discussed in the text.
(a) pseudorapidity dependence for central events
(PCal $E_t$ energy region 310--330~GeV);
(b) PCal $E_t$ dependence of $E_t$ per charged particle
in the region $1.5<\eta<2.0$.

\end{enumerate}

%
%
%
%
%

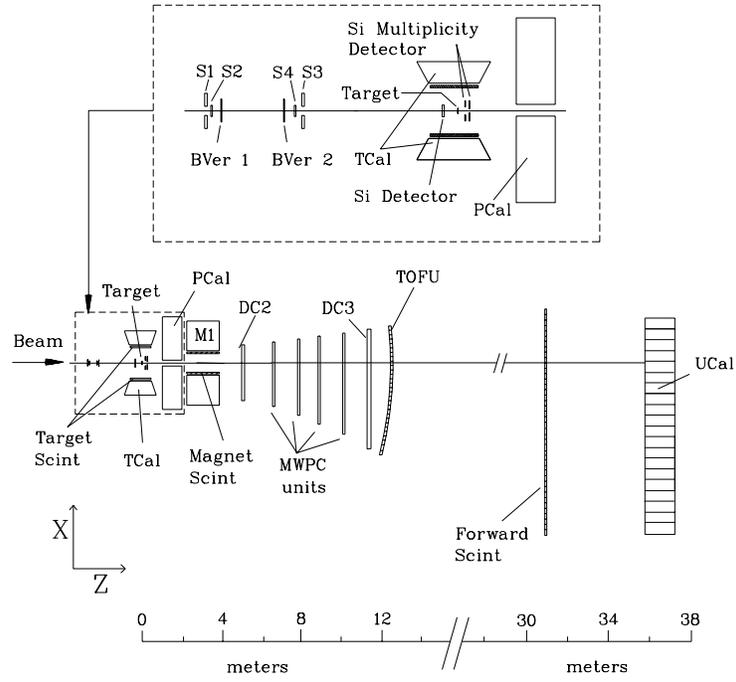
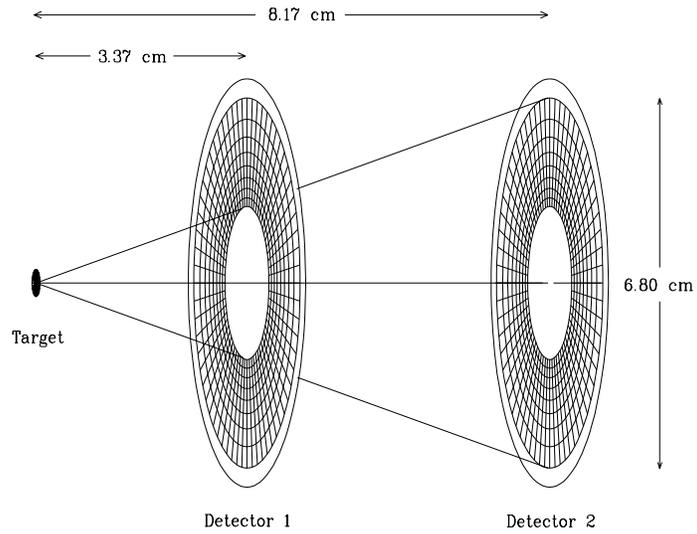
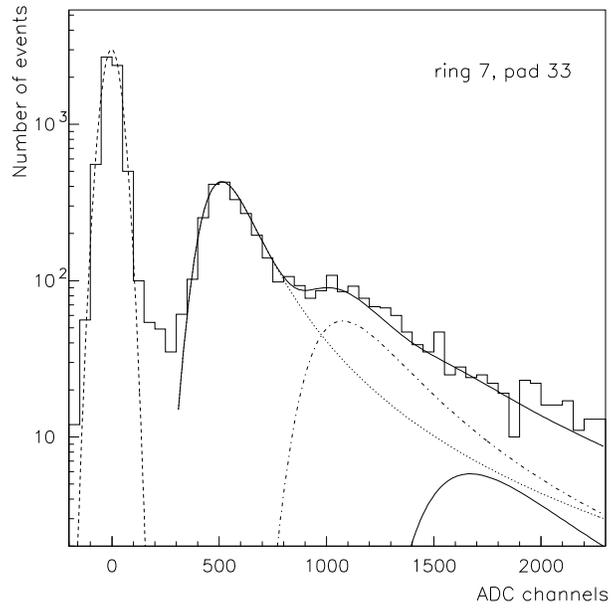
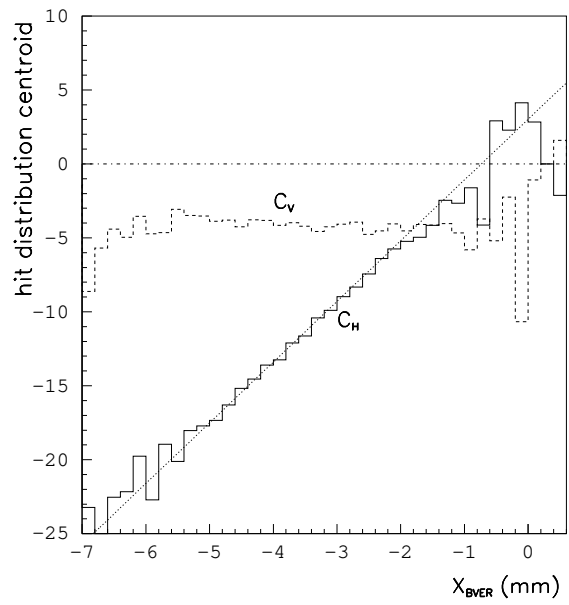
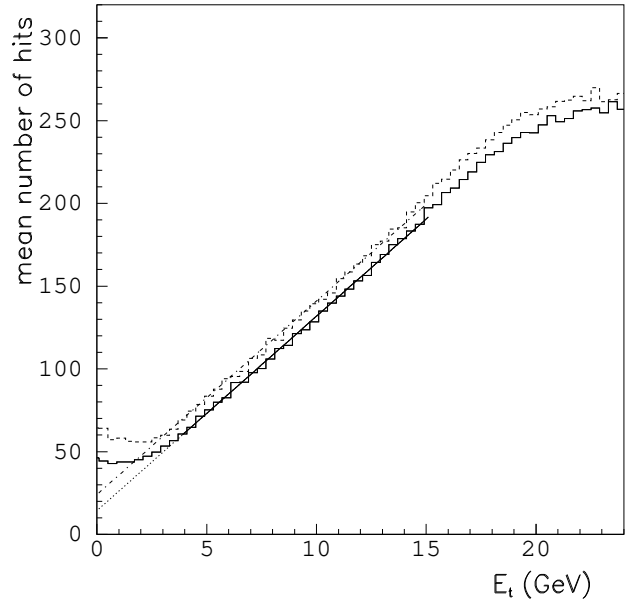
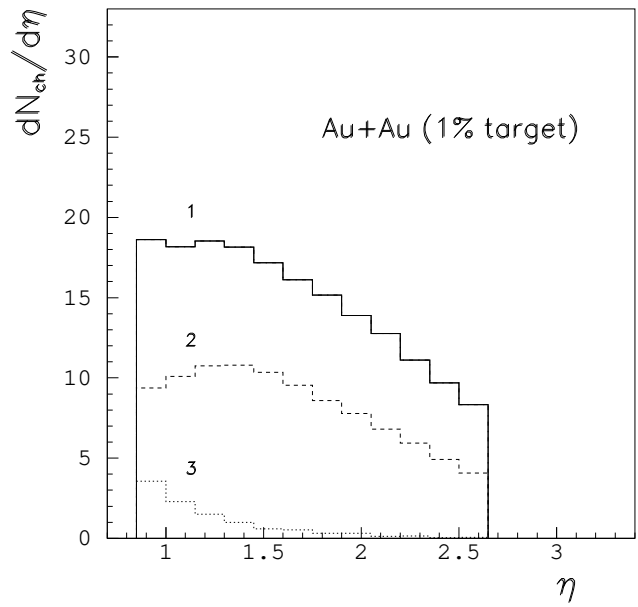
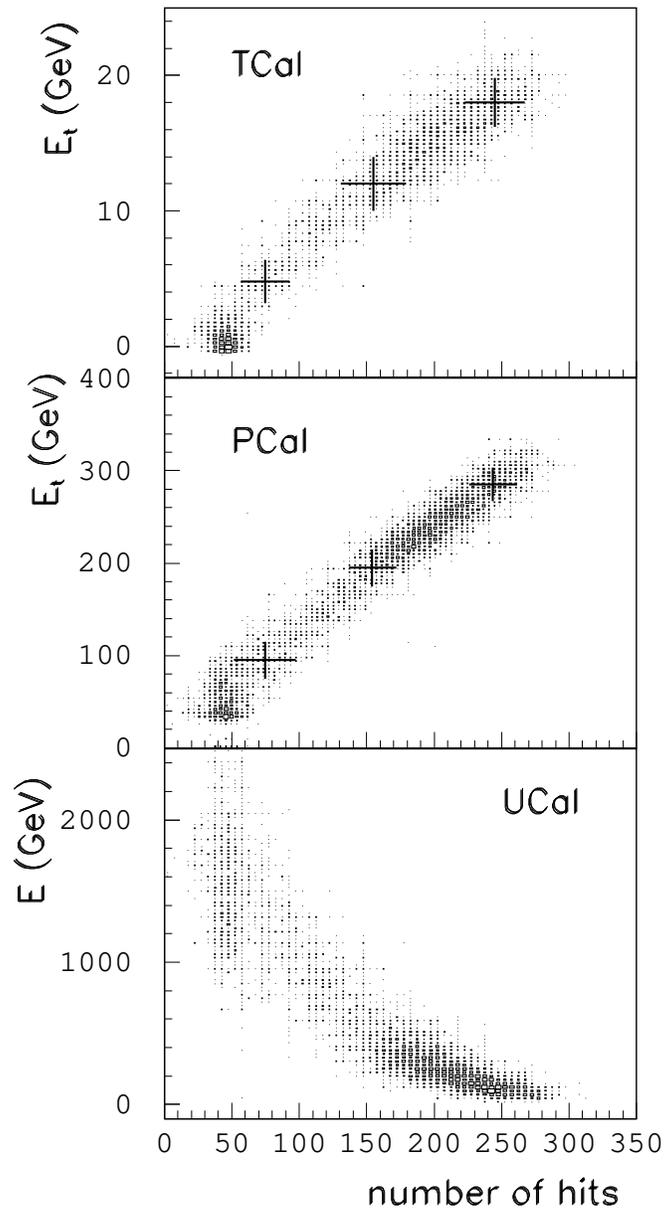
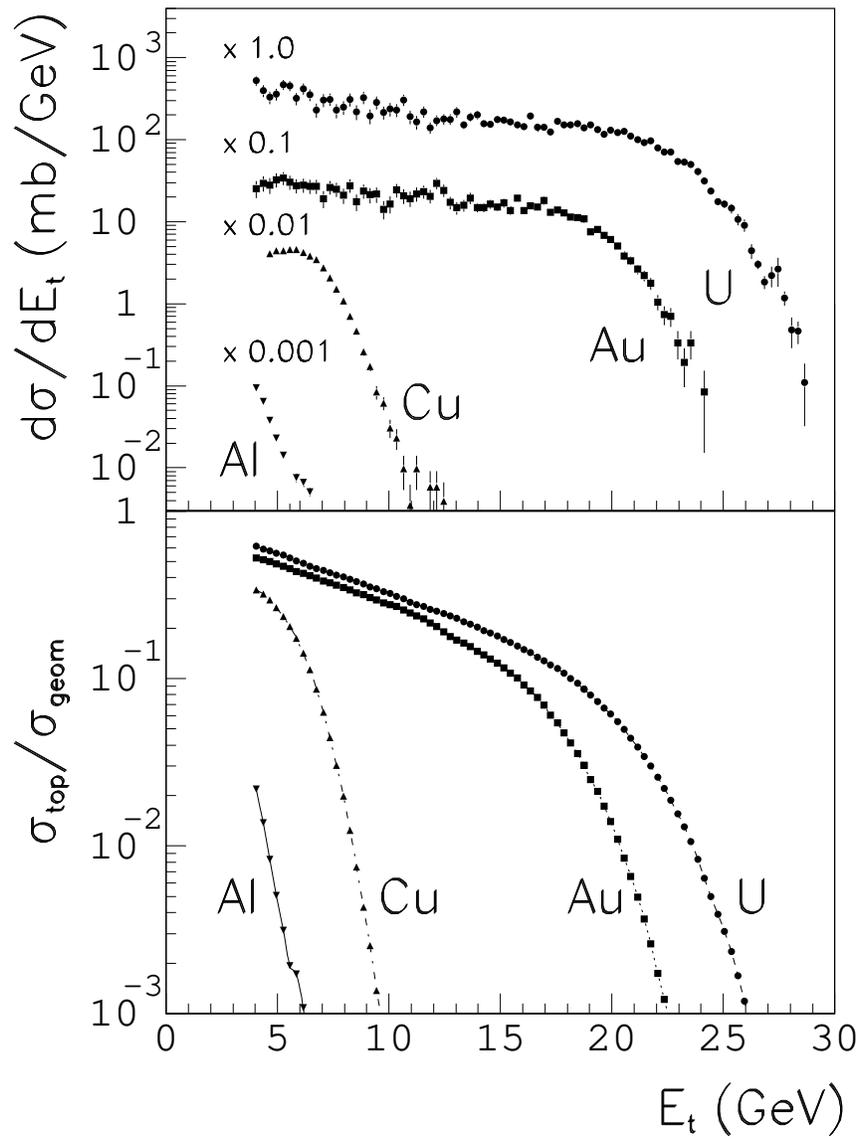
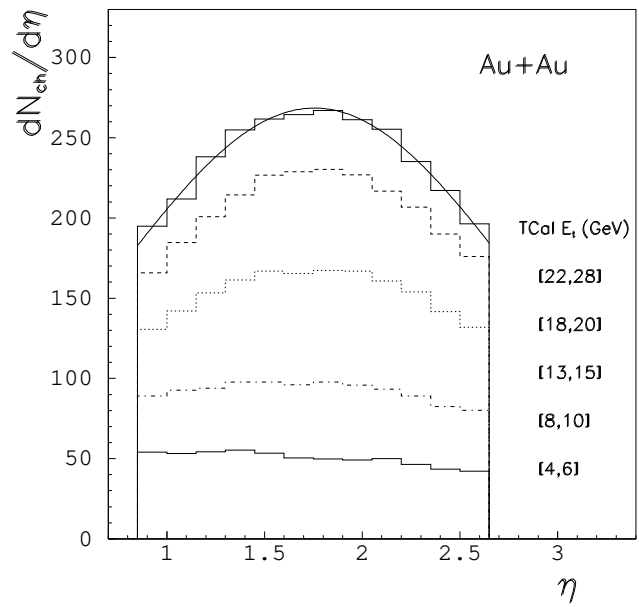
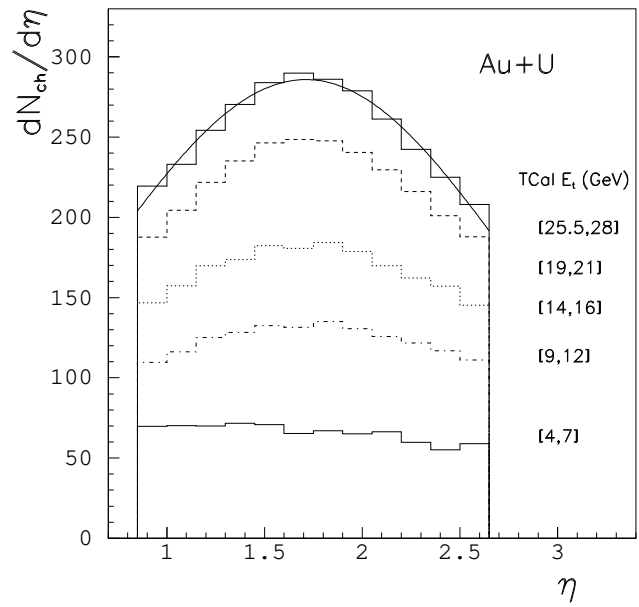
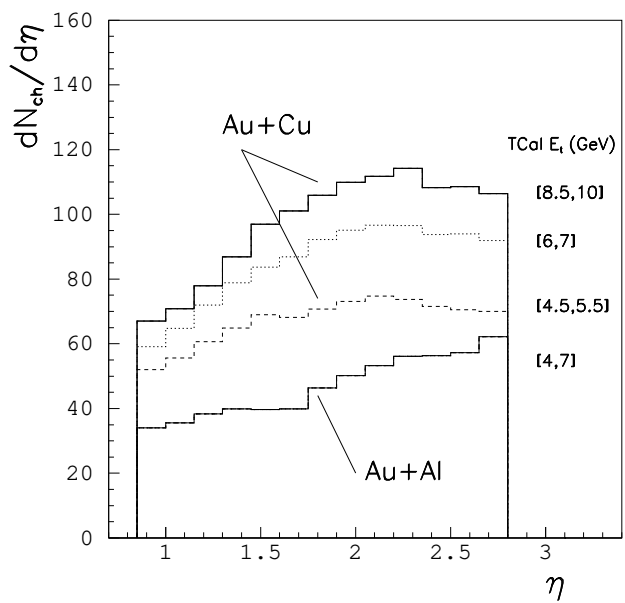
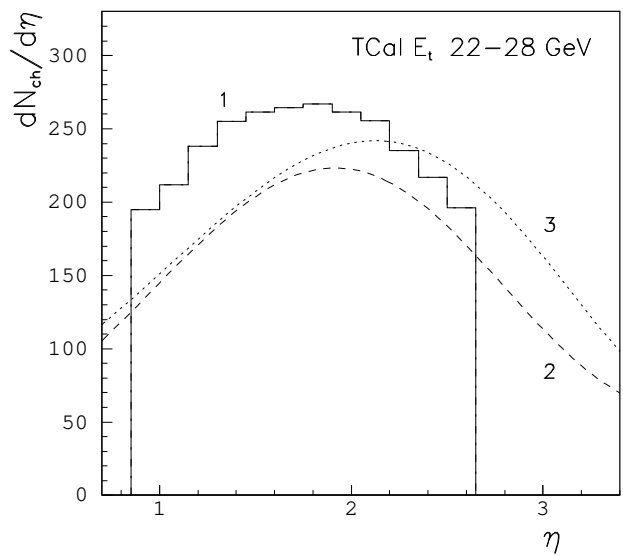
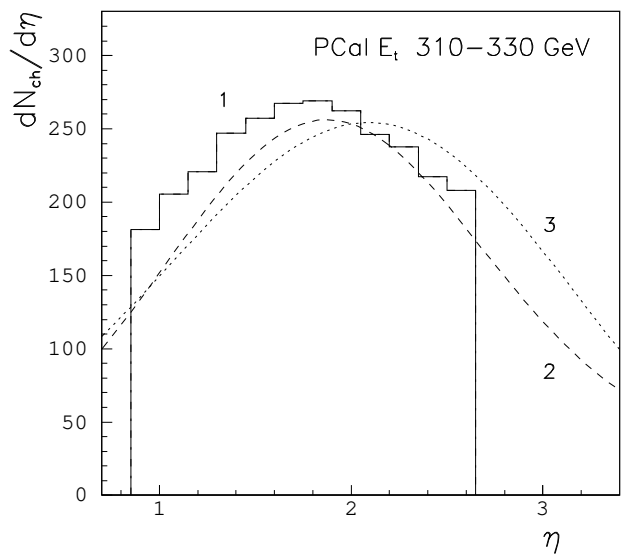
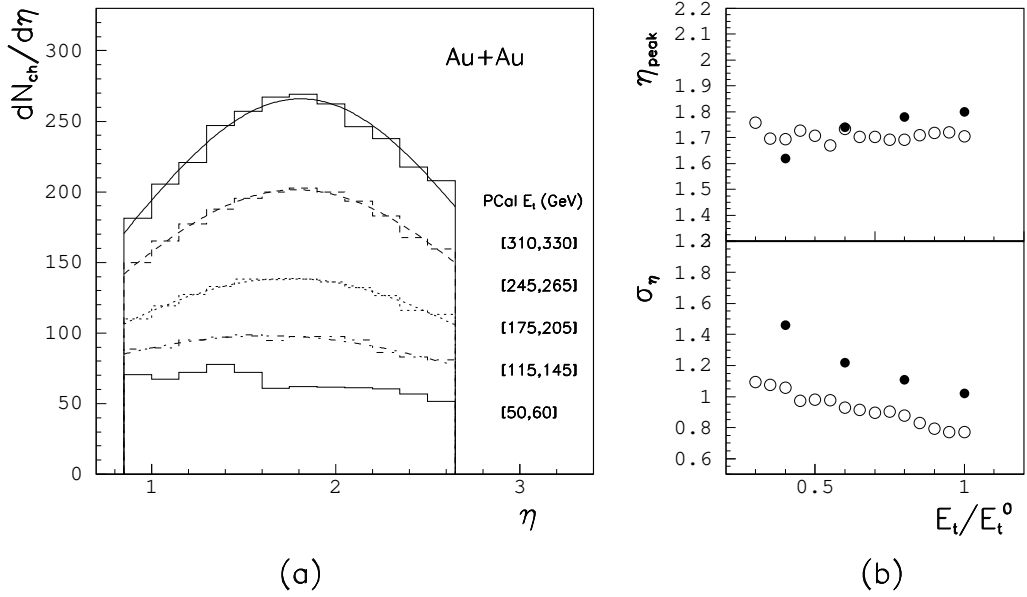
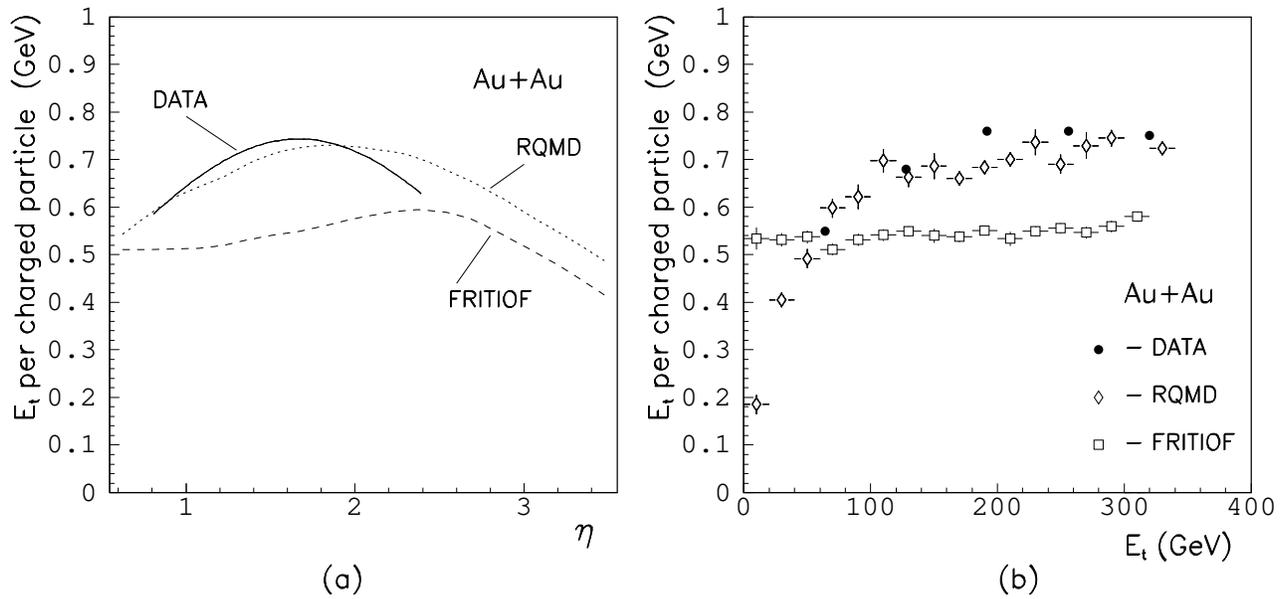
\end{document}